\pdfoutput=1
\documentclass[11pt]{article}

\usepackage{csquotes}

\usepackage[round,longnamesfirst]{natbib}
\usepackage{amsmath, amsthm, amssymb, amsbsy,enumitem}
\setitemize{noitemsep}
\usepackage{array}
\usepackage{graphicx}
\usepackage{multirow}
\usepackage{cancel}
\usepackage[noend]{algorithmic}
\usepackage{algorithm}
\usepackage{setspace}
\usepackage{url}
\usepackage{verbatim}

\usepackage[bookmarksnumbered=true,backref=page,hyperfigures=true]{hyperref}

\usepackage{ifthen}
\usepackage{color}

\def\y{{\boldsymbol{y}}}
\def\Y{{\boldsymbol{Y}}}

\def\X{{\boldsymbol{X}}}

\def\bpi{{\boldsymbol{\pi}}}

\def\natmap{{\boldsymbol{\eta}}}
\def\curvpar{{\boldsymbol{\theta}}}
\def\disspar{{\boldsymbol{\omega}}}
\def\genstat{{\boldsymbol{g}}}

\def\nactors{n}
\def\actors{N}

\def\form{^+}
\def\diss{^-}

\def\changeij{{\boldsymbol{\Delta}\sij}}
\def\setsub{\backslash}

\def\age{a}

\DeclareMathOperator{\logit}{logit}

\DeclareMathOperator{\Geometric}{Geometric}

\def\dysY{\mathbb{Y}}
\def\netsY{\mathcal{Y}}

\DeclareMathOperator{\Prob}{Pr}
\def\Peg{\Prob_{\natmap,\genstat}}
\def\PegF{\Prob_{\natmap\form,\genstat\form}}
\def\PegD{\Prob_{\natmap\diss,\genstat\diss}}

\def\cegF{c_{\natmap\form,\genstat\form}}
\def\cegD{c_{\natmap\diss,\genstat\diss}}
\def\netsYF{\mathcal{Y}\form}
\def\netsYD{\mathcal{Y}\diss}
\def\YF{\Y\form}
\def\YD{\Y\diss}
\def\yF{\y\form}
\def\yD{\y\diss}
\def\genstatF{\genstat\form}
\def\genstatD{\genstat\diss}
\def\curvparF{\curvpar\form}
\def\curvparD{\curvpar\diss}
\def\natmapF{\natmap\form}
\def\natmapD{\natmap\diss}

\DeclareMathOperator{\ilogit}{logit^{-1}}

\def\NN{\mathbb{N}}
\def\pij{(i,j)}

\def\ijdysY{\pij\in\dysY}
\def\iactors{i\in\actors}

\def\sij{_{i,j}}

\def\Yij{\Y\sij}
\def\yij{\y\sij}

\def\natpar{\natmap(\curvpar)}
\def\natparF{\natmap\form(\curvpar\form)}
\def\natparD{\natmap\diss(\curvpar\diss)}
\def\curvparF{\curvpar\form}
\def\curvparD{\curvpar\diss}

\newcommand{\myexp}[1]{\exp\left(#1\right)}

\newcommand{\I}[1]{1_{#1}}
\newcommand{\pkg}[1]{\textbf{#1}}
\newcommand{\proglang}[1]{\textsf{#1}}

\newcommand{\yat}[1]{\y^{t#1}}
\newcommand{\Yat}[1]{\Y^{t#1}}
\newcommand{\Yyat}[1]{\Yat{#1}=\yat{#1}}

\newcommand{\yatij}[1]{\y^{t#1}\sij}
\newcommand{\Yatij}[1]{\Y^{t#1}\sij}

\providecommand{\abs}[1]{\left\lvert#1\right\rvert}

\graphicspath{{graphics/}{common/}}

\title{Modeling Tie Duration in ERGM-Based Dynamic Network Models}

\author{Pavel N. Krivitsky\\\texttt{p.krivitsky@unsw.edu.au}\\School of Mathematics and Statistics\\University of New South Wales\\Sydney, NSW, Australia}

\date{April 2012 (original)\\
\today{} (repost)}

\begin{document}
\maketitle
\begin{quotation}
  \small
  \noindent\begin{center}\textbf{History}\end{center}
  
  This preprint was originally published to Penn State University Department of Statistics web site as \emph{Technical Report 12--02} in April 2012. It was subsequently lost, along with others, in a web site migration. In order to return it to the public record, we are reposting it, unmodified except as noted here:
  \begin{itemize}
  \item Penn State Statistics technical report title page has been replaced.
  \item Baseline font size has been enlarged for readability.
  \item Author affiliation and contact information has been added.
  \item Some items in the bibliography have been reformatted or updated.
  \item These changes may affect pagination.
  \end{itemize}
  This version may be superseded by other versions in the future.
\end{quotation}

\clearpage\begin{abstract}
  \citet{KrHa14s} proposed a Separable Temporal ERGM (STERGM)
  framework for modeling social networks, which facilitates separable
  modeling of the tie duration distributions and the structural
  dynamics of tie formation. In this note, we explore the hazard
  structures achievable in this framework, with first- and
  higher-order Markov assumptions, and propose ways to model a variety
  of duration distributions in this framework.
\end{abstract}

\section{Introduction}

Modeling of dynamic networks --- networks that evolve over time ---
has applications in many fields, particularly epidemiology and social
sciences. Exponential-family random graph ($p^*$) models (ERGMs) for social
networks are a natural way to represent dependencies in
cross-sectional graphs and dependencies between graphs over time,
particularly in a discrete context, and \citet{robins2001rgm} first
described this approach. \citet{hanneke2010dtm} also define and
describe what they call a Temporal ERGM (TERGM), postulating an
exponential family for the transition probability from a network at
time $t$ to a network at time $t+1$.

\citet{holland1977dms}, \citet{frank1991sac}, and others describe
\emph{continuous-time} Markov models for evolution of social networks
\citep{doreian1997esn}, and the most popular parametrization is the
\emph{actor-oriented} model described by \citet{snijders2005mln},
which can be viewed in terms of actors making decisions to make and
withdraw ties to other actors.

Arguing that \enquote{social processes and factors that result in ties
  being formed are not the same as those that result in ties being
  dissolved}, \citet{KrHa14s} introduced a separable
formulation of discrete-time models for network evolution parametrized
in terms of a process that controls formation of new ties and a
process that controls dissolution of extant ties, in which both
processes are (possibly different ERGMs), calling them \emph{Separable
  Temporal ERGMs} (STERGMs). Thus, the model separates the factors
that affect \emph{incidence} of ties --- the rate at which new ties
are formed --- from their \emph{duration} --- how long they tend to
last once they do. This latter aspect, combined with its discrete-time
nature, in turn, allows straightforward modeling of complex tie hazard
structures and duration distributions. In this work, we discuss how a
variety of these can be modeled.

In Section~\ref{sec:sep}, we review the STERGM framework. In
Section~\ref{sec:markov-1} we discuss tie hazard structures that can
be induced in the framework under the first-order Markov assumption
--- that the transition probability does not take into account
duration explicitly, while in Section~\ref{sec:hazard}, we propose a
variety of ways to model tie hazard explicitly.
 
\section{\label{sec:sep}Separable temporal ERGM}
We now review the model proposed by \citet{KrHa14s} and
define some additional notation. The following overview borrows
heavily from \citet{krivitsky2012mdn}. Using their notation, let
$\actors$ be the set of $\nactors=\abs{\actors}$ actors of interest,
labeled $1,\dotsc,\nactors$, and let
$\dysY\subseteq\actors\times\actors$ be the set of dyads (potential
ties) among the actors, with $\ijdysY$ directed if modeling directed
relations and $\{i,j\}\in\dysY$ for undirected networks. $\dysY$ may
be a proper subset: for example, self-loops with $i=j$ are often
excluded. Then, the set of possible networks $\netsY$ is the power set
of dyads, $2^\dysY$. For a network at time $t-1$, $\yat{-1}$,
\citet{KrHa14s} define $\netsYF(\yat{-1})= \{\y\in
2^\dysY:\y\supseteq\yat{-1}\}$ be the set of networks that can be
constructed by forming zero or more ties in $\yat{-1}$ and
$\netsYD(\yat{-1})= \{\y\in 2^\dysY:\y\subseteq\yat{-1}\}$ be the set
of networks that can be constructed by dissolving zero or more ties in
$\yat{-1}$.

Given $\yat{-1}$, the network $\Yat{}$ at time $t$ is modeled as a
consequence of some ties being formed according to a conditional ERGM
\begin{equation*}\PegF(\YF = \yF|\Yyat{-1};\curvparF )=\frac{\myexp{\natparF\cdot \genstatF(\yF,\yat{-1})}}{\cegF(\curvparF ,\yat{-1})},\ \yF\in\netsYF(\yat{-1})\end{equation*}
specified by model parameters $\curvparF$, sufficient statistic
$\genstatF$, and, optionally, a canonical mapping $\natmapF$; and some
dissolved according to a conditional ERGM
\begin{equation*}\PegD(\YD = \yD|\Yyat{-1};\curvparD )=\frac{\myexp{\natparD\cdot \genstatD(\yD,\yat{-1})}}{\cegD(\curvparD ,\yat{-1})},\ \yD\in\netsYD(\yat{-1}),\end{equation*} 
specified by (usually different) $\curvparD$, $\genstatD$, and
$\natmapD$. Their normalizing constants $\cegF(\curvparF ,\yat{-1})$
and $\cegD(\curvparD ,\yat{-1})$ sum their respective model kernels
over $\netsYF(\yat{-1})$ and $\netsYD(\yat{-1})$, respectively.
$\Yat{}$ is then evaluated by applying the changes in $\YF$ and $\YD$ to $\yat{-1}$: $\Yat{}=\yat{-1}\cup (\yF\setsub\yat{-1})\setsub
(\yat{-1}\setsub\yD)=\yF\setsub(\yat{-1}\setsub\yD)=\yD\cup(\yF\setsub\yat{-1})$.

Although an ERGM is a model for a whole network, many ERGM sufficient
statistics have a local interpretation in the form of \emph{change
  statistics}
$\changeij\genstat(\y)=\genstat(\y\cup\{\pij\})-\genstat(\y\setsub\{\pij\})$,
the effect that a single dyad $\pij$ has on the model's sufficient
statistic and thus on its conditional probability given the rest of
the network. \citep{hunter2008epf} For models with dyadic
independence, the conditional probability is the same as the marginal
probability, so
$\Peg(\Yij=1)=\ilogit(\natpar\cdot\changeij\genstat(\y))$. When
applied to the dissolution phase, this is the probability of an extant
tie being preserved during a given time step.

When discussing the tie hazard structure of a model, we define
$\age(\yatij{})$, the \emph{age} of a network tie $\pij$ at time $t$
that is present at time $t$, to be the number of time steps that had
elapsed since the tie was formed, as of time $t$. This is in contrast
to a tie's \emph{duration}, which is a measure of how long a tie
ultimately lasts, with the distinction being analogous to that between
a person's age in a given year and their ultimate lifespan. In a
STERGM, a tie cannot be formed and dissolved in the same time step, so
$\age(\yij)\ge 1$. Notably $\age(\yatij{})$ is, implicitly, a function
of $\yatij{-1}$, $\yatij{-2}$, etc., up to
$\yatij{-(\age(\yatij{})+1)}$, at which point it becomes known how
long ago the tie was formed.

\section{\label{sec:markov-1}{Tie hazards for first-order Markov models}}
We begin by considering hazard properties of first-order Markov
models: models where a network $\Yat{}$ only depends on networks
$\Yat{-q}$, $q>1$, through $\Yat{-1}$.

\subsection{Constant hazard}
When only dyad-independent, implicitly dynamic dissolution statistics
--- statistics that only depend on $\yat{-1}$ through $\yD$ --- are
used, such as edge counts, mixing counts, and actor and edge
covariates, each dyad has a geometric (discrete memoryless)
distribution, although depending on the statistics used and exogenous
covariates, each dyad may have a different expected
duration. \citep{KrHa14s} Being memoryless, the geometric
distribution has a constant hazard function:
\begin{align*}
  h_{\Geometric(p)}(x)&=\frac{f_{\Geometric(p)}(x)}{1-F_{\Geometric(p)}(x-1)}
=p.
\end{align*}
This is the case described and applied by \citet{krivitsky2012mdn}.

\subsection{\label{sec:hazard-dyad-dep}Non-constant hazard through dyadic dependence}
When dyadic dependence is introduced into the dissolution process, the
marginal hazard function of each dyad may no longer be constant. For
example, if the formation model \enquote{enforces} monogamy by
\enquote{encouraging} formation of an actor's first tie and
\enquote{penalizing} the formation of the second tie, while the
dissolution model has a statistic that reduces the dissolution hazard
of ties when they are monogamous, say,
\[
  \genstat\form(\yF)=\left(\abs{\yF},\sum_{\iactors} \I{\abs{\yF_i}=1}\right),\quad
  \genstat\diss(\yD)=\left(\abs{\yD},\sum_{\iactors} \I{\abs{\yD_i}=1}\right),
\]
with $\curvparF=(-,+)$ and $\curvparD=(+,+)$ --- negative coefficient
on formation phase edge counts and positive coefficient on dissolution
edge count (to produce relatively slow network evolution), and
positive coefficients on the counts of actors with degree 1.

The dissolution phase is a draw from a dyad-dependent ERGM, so
deriving the exact hazard function for this model, even conditional on
$\yat{-1}$, is intractable, but heuristically, a tie may be found in
one of two scenarios:
\begin{enumerate}[partopsep=0pt,topsep=0pt,parsep=0pt,itemsep=0pt]
\item It is the only tie incident on the actors on which it is incident (i.e. actors $i$ and $j$ are both isolates without it).\label{enum:onlytie}
\item It is not the only tie incident on the actors on which it is
  incident (i.e. either $i$ or $j$ has other
  ties).\label{enum:notonlytie}
\end{enumerate}
The positive coefficient $\curvparD_2$ increases the hazard of those
ties which are not their actors' only ties (ties in
Scenario~\ref{enum:notonlytie}) so ties in
Scenario~\ref{enum:notonlytie} would have a relatively high hazard,
with dissolution likely until only one tie is left. However, when only
one tie is left (i.e. Scenario~\ref{enum:onlytie}), $\curvparD_2$
reduces the hazard of that tie \emph{and} the positive coefficient
$\curvparF_2$ reduces the probability that another tie incident on
either of the actors will be formed during a given time step.

This means that a new tie that does form between actors which already
have ties will have a relatively high hazard, but so will other ties
incident on those actors, and if the new tie is the
\enquote{survivor}, its hazard will decrease and monogamy bias in
formation phase (if present) will reduce any \enquote{competition} it
might face. This results in a hazard function which is high at first
and decreases over time. (A new tie that forms between actors that do
not already have any ties will have a constant hazard.)
  
To illustrate this, we conduct a simulation of a dynamic network with
50 actors. We used \proglang{R} package \pkg{ergm}
\citep{hunter2008epf,handcock2012epf} to simulate four runs of the
network process, each 11,000 time steps, with the following parameter
configurations:\label{loc:dyad-dep-haz-par}
\begin{description}[partopsep=0pt,topsep=0pt,parsep=0pt,itemsep=0pt]
\item[$\curvpar_{\text{I}}=(\curvparF,\curvparD)=(-6,0,2,0)$] total dyadic
  independence;
\item[$\curvpar_{\text{D}}=(-6,0,2,2)$] formation
  dyadic-independent, dissolution dyad-dependent;
\item[$\curvpar_{\text{F}}=(-6,2,2,0)$] formation
  dyad-dependent, dissolution dyadic-independent; and
\item[$\curvpar_{\text{B}}=(-6,2,2,2)$] both formation and dissolution
  dyad-dependent.
\end{description}
Since the goal of this simulation is to contrast the differences in
tie hazard functions due to monogamy bias parameters, these
configurations all create short-duration dynamic network
processes. The four configurations produce networks with different
(equilibrium) degree distributions and densities, but the quantity of
interest is the edgewise hazard function, which is effectively
adjusted for the number of edges in the network.
\begin{figure}
  {\noindent
    \begin{center}
      \includegraphics[width=.8\columnwidth,keepaspectratio]{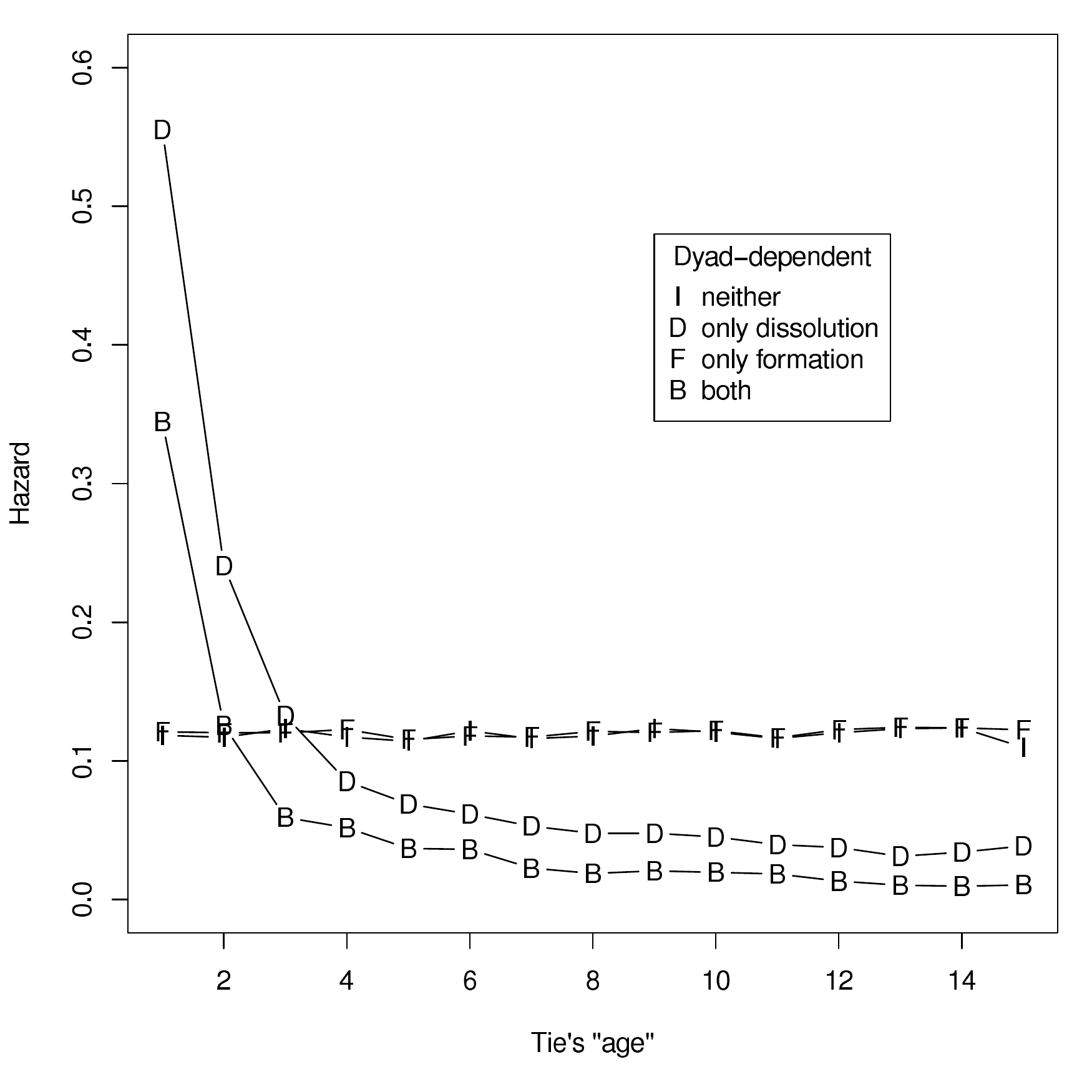}
    \end{center}
  }   
  \caption[Tie hazards under models with and without dyadic
  independence]{\label{fig:dyad-dep-haz}Estimated tie hazards under
    four parameter configurations given on
    page~\pageref{loc:dyad-dep-haz-par}. Note that hazards at
    non-integral ages are meaningless, so the lines between data
    points are only drawn to make the series easier to follow. Note
    that constant hazard corresponds to geometric duration
    distribution.}
\end{figure}
\begin{table}
  \caption[Equilibrium statistics for hazard functions
  dyad-dependence example]{\label{tab:dyad-dep-haz}Simulated equilibrium statistics under
    the four parameter configurations given on
    page~\pageref{loc:dyad-dep-haz-par}.}
  
  {\noindent
    \begin{center}
      \begin{tabular}{|c|cc|}
        \hline
        Parameter& Network & Prop. of actors \\
        configuration & density & with 1 tie  \\
        \hline
        I &0.020  & 0.37 \\
        D &0.022  & 0.73 \\
        F &0.019  & 0.74 \\
        B &0.020  & 0.96 \\
        \hline
      \end{tabular}
    \end{center}
  }
\end{table}

We estimate the discrete hazard function
\begin{align*}
  \widehat{\Prob}(X=x|X\ge x)&= \frac{\text{\# ties terminated with duration $x$}}{\text{\# ties terminated with duration of at least $x$}}
\end{align*}
for tie ages $x=1,\dotsc,15$, for each parameter
configuration. Durations of ties formed in the first 1,000 time steps
were excluded as burn-in.  The results from the simulation are given
in Figure~\ref{fig:dyad-dep-haz} and Table~\ref{tab:dyad-dep-haz}.  As
expected, under temporal dyadic independence ($\curvpar_{\text{I}}$),
the hazard is constant --- it is $1-\ilogit(2)\approx 0.119$ --- and
dyadic dependence limited to tie formation ($\curvpar_{\text{F}}$)
does not change this. When dissolution is dyad-dependent (both
$\curvpar_{\text{B}}$ and $\curvpar_{\text{D}}$), the hazard is
initially high, but then declines, as expected. However, it declines
to a slightly lower level when both formation and dissolution have a
monogamy bias ($\curvpar_{\text{B}}$): a monogamous tie not only has
its hazard reduced in the dissolution but prevents any
hazard-increasing \enquote{competitors} from arising in
formation. This is not the case when formation is dyad-independent
($\curvpar_{\text{D}}$), and a monogamous tie is always potentially
subject to this \enquote{competition}.
  
Thus, non-constant dyad hazards can be induced by dyadic dependence in
dissolution, and if thus induced, they may be affected by dependence
in formation as well. The hazard of a given dyad during a given time
step is a function only of the state of the network at the beginning
of that time step, so even though the hazards are not constant, the
Markov property of the process is preserved. In the following
sections, we relax this, and describe explicit non-constant hazards.

\section{\label{sec:hazard}Higher-order Markov specifications}
One of the advantages discrete-time models have over continuous-time
models is simpler control over the duration distribution. In the
context of STERGMs and TERGMs in general,
this is done by manipulating edgewise hazard functions. In this
section, we describe several ways in which non-memoryless tie duration
distributions can be induced.

\subsection{Piecewise-constant hazard model}
\begin{figure}
  {\noindent
    \begin{center}
      \begin{tabular}{cc}
        \includegraphics[width=0.48\columnwidth,keepaspectratio]{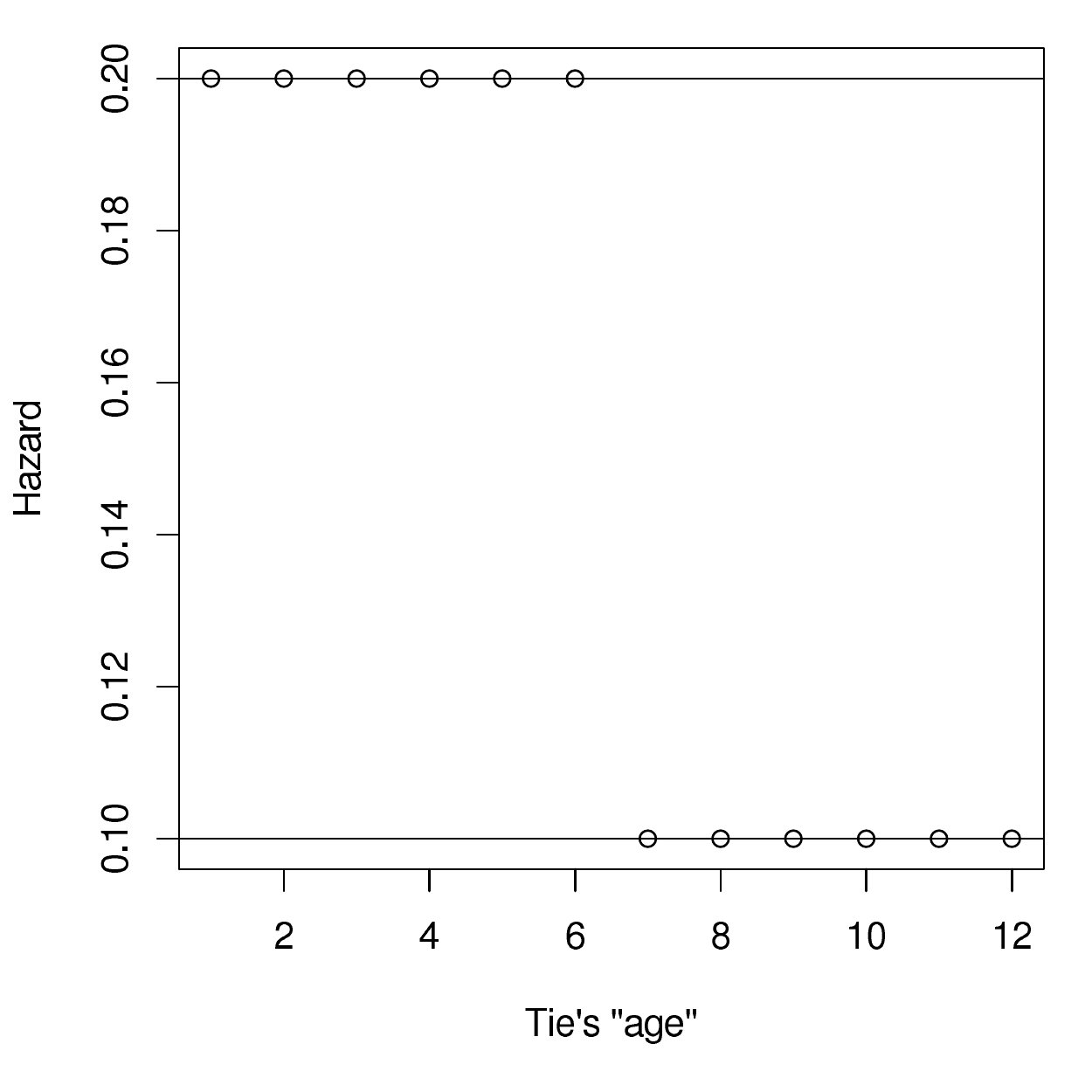}&
        \includegraphics[width=0.48\columnwidth,keepaspectratio]{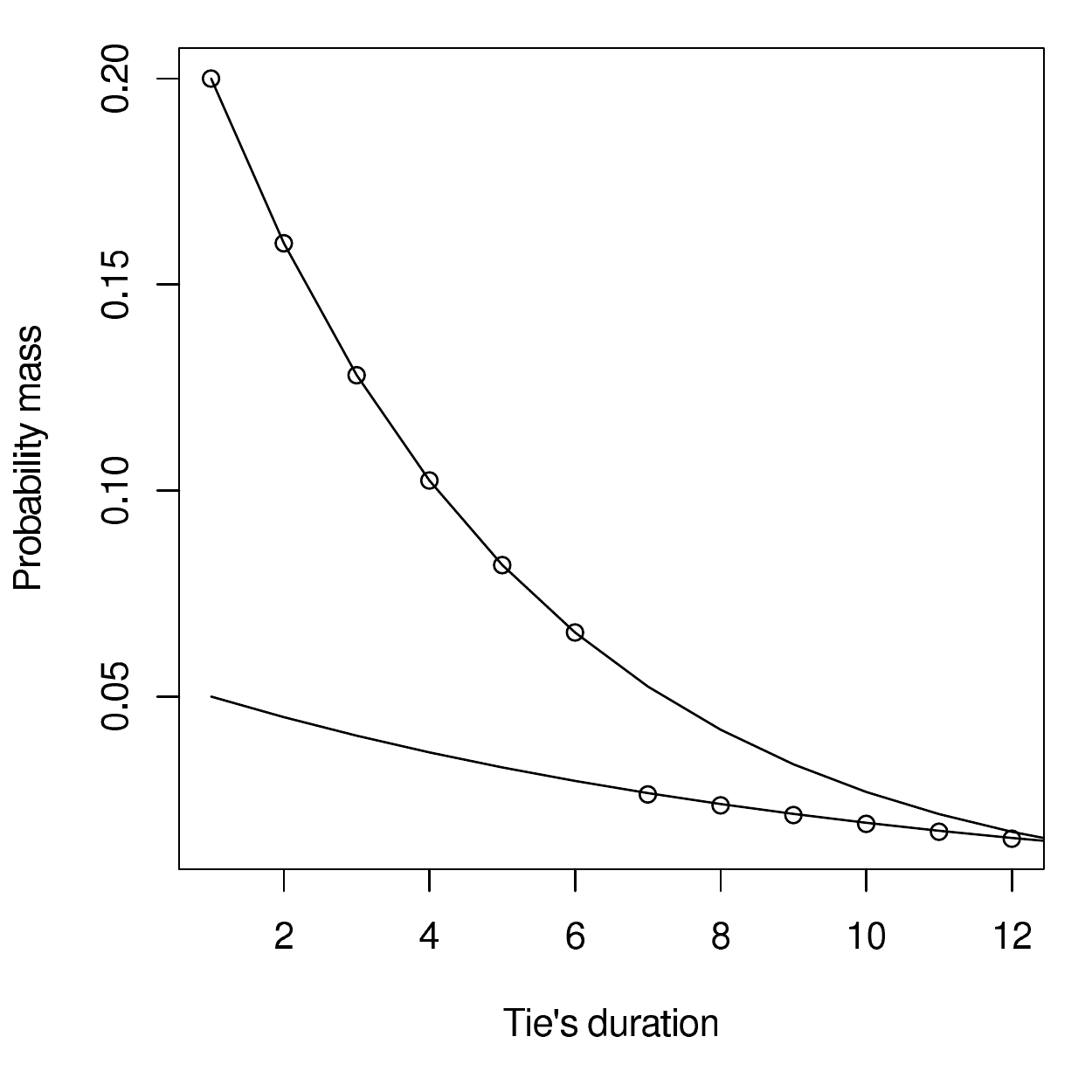}\\
        (a) hazard function&
        (b) probability mass function  \\
      \end{tabular}
    \end{center}
  }   
  \caption{\label{fig:piecewise-hazard} Piecewise-constant hazard function and resulting duration distribution.}
\end{figure}

The simplest way to directly induce non-constant tie hazard is by
modifying it by a fixed value for some set of age values. For example,
let
\[\genstat\diss(\yD)=\left(\abs{\yD},\sum_{\pij \in \yD} \I{\age(\yD\sij)\in A}\right),\]
for some set $A\subset \NN$. $\genstat\diss_2$ counts the
number of ties in the network that whose age at the time point of
interest is in set $A$. (For most practical purposes, $A$ is a discrete
interval.) With change statistic,
\[\changeij\genstat(\yat{},\yat{-1})=\left(\yatij{-1},\yatij{-1}\I{\age(\yD\sij)\in A}\right),\]
leads to the probability of a tie being preserved in a time step of
\[\Peg(\Yatij{}=1|\Yatij{-1}=1;\curvparD)=\ilogit\left(\curvparD_1+\curvparD_2 \I{\age(\yD\sij)\in A}\right),\]
and results in the probability of it being dissolved (the hazard)
\[h(\age(\yD\sij))=\Peg(\Yatij{}=0|\Yatij{-1}=1;\curvparD)=\ilogit\left(-\curvparD_1-\curvparD_2  \I{\age(\yD\sij)\in A}\right).\] 
In this case, the hazard can attain
two values, and if $A=\{1,\dotsc,a_0\}$ for some $a_0$, the duration
distribution, which can be computed recursively from the hazard
function $h$, as follows:
\[
  f(1)=h(1),\quad
  f(x)=h(x)\left(1-\sum_{i=1}^{x-1}f(i)\right)
\]
giving duration distribution
\[f(x)=\begin{cases}
  \left(\ilogit\left(\curvparD_1+\curvparD_2 \right)\right)^{x-1}\ilogit\left(-\curvparD_1-\curvparD_2 \right) & \text{for $x\le a_0$}\\
  \left(\ilogit\left(\curvparD_1+\curvparD_2 \right)\right)^{a_0}\left(\ilogit\left(\curvparD_1 \right)\right)^{x-a_0-1} \ilogit\left(-\curvparD_1\right) & \text{for $x> a_0$}
\end{cases},\]
shown in Figure~\ref{fig:piecewise-hazard} for $a_0=6$, $\curvparD_1=\logit(0.9)$ (hazard of $0.1$) and $\curvparD_2=\logit(0.8)-\curvparD_1$ (hazard of $0.2$).

If $A$ is finite, with $a_0=\sup(A)$, this model for network evolution
$(a_0+1)$th-order Markov: beyond $a_0$, the hazard reverts to the
baseline $\ilogit(-\curvparD_1)$, regardless of the state of networks
prior to $\yat{-a_0-1}$.

It is straightforward to extend this formulation to more hazard
levels.

\subsubsection{Application to formation}
Rather than viewing age as an attribute of a tie, we can view it as
an attribute of any dyad --- the time since the last toggle in
\emph{either} direction --- and in the formation phase, it can be used
to, for example, penalize reformation of recently-dissolved ties.

\subsection{\label{sec:finite-mix-haz}Finite mixture model for duration}
Another possible source of non-constant hazard in the duration
distribution is unobserved (latent) classes of ties. An example of
this is models for networks of sexual partnerships, which may be
short-term or long-term. Dyad-dependent but (temporally) Markovian
features of the model, such as a monogamy bias for dissolution
demonstrated in Section~\ref{sec:hazard-dyad-dep}, can account for
some of this. Alternatively, the duration distribution can be modeled
as a finite mixture of simpler distributions: let there be $m$ latent
types of relationships, indexed $1,\dotsc,m$, $\X_1,\dotsc,\X_m$ be
the duration distributions of different relationship types
parametrized by $\disspar$, let and $\bpi_1,\dotsc,\bpi_m$,
$\bpi_k>0$, $\sum_{k=1}^m\bpi_k=1$, be their \emph{incidence}. That
is, at the time a tie forms, the probability that it is a tie of the
type with duration distributed as $\X_k$ is $\bpi_k$. This is
different from the tie class \emph{prevalence} in the population,
since that is also a function of duration of ties (that differs
between types): the tie types with higher expected duration will be
disproportionately more prevalent relative to their incidence. Let $X$
be the marginal duration distribution of a tie. For notational
convenience, let $\curvparD=(\disspar,\bpi)$.

\begin{figure}
  {\noindent
    \begin{center}
      \begin{tabular}{cc}
        \includegraphics[width=0.48\columnwidth,keepaspectratio]{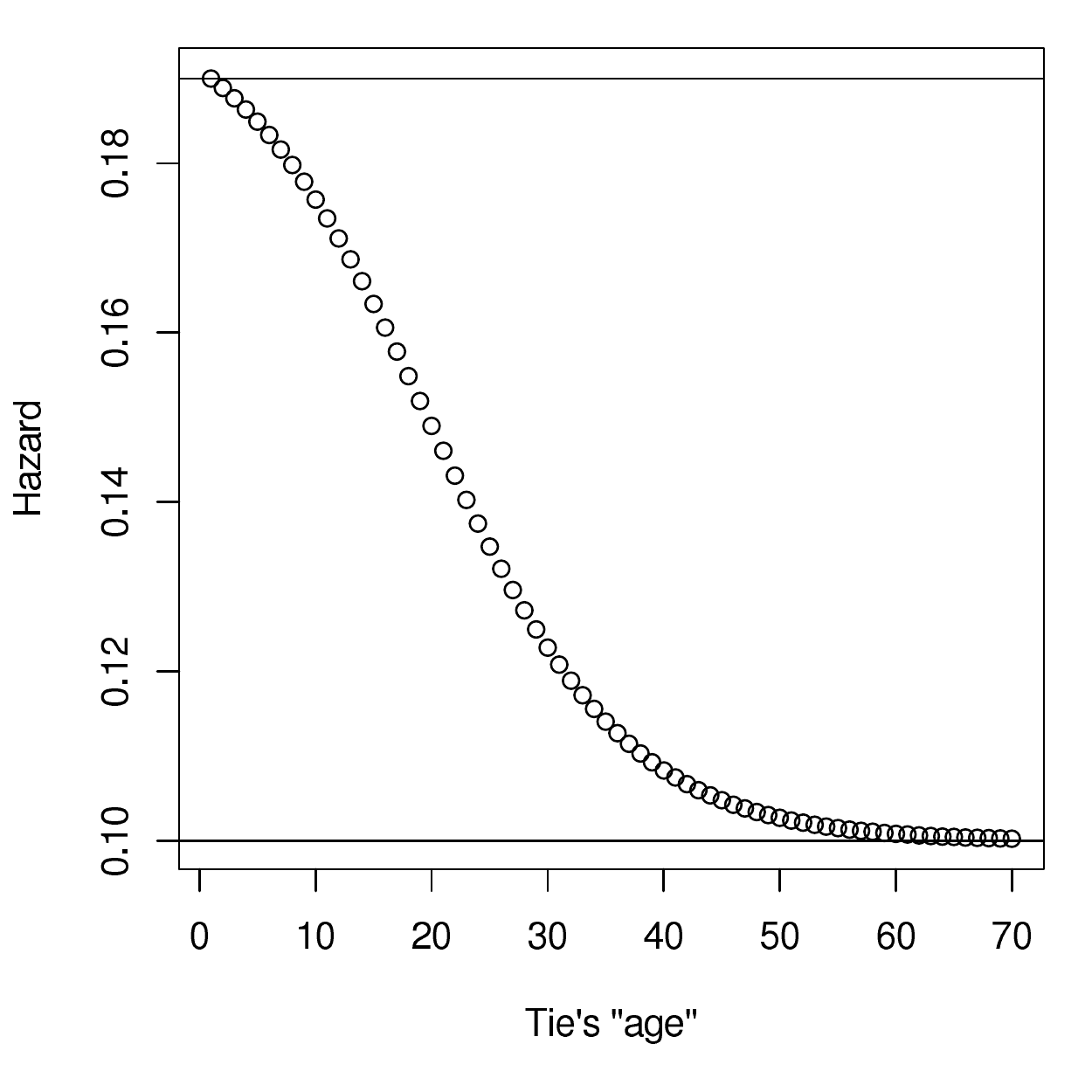}&
        \includegraphics[width=0.48\columnwidth,keepaspectratio]{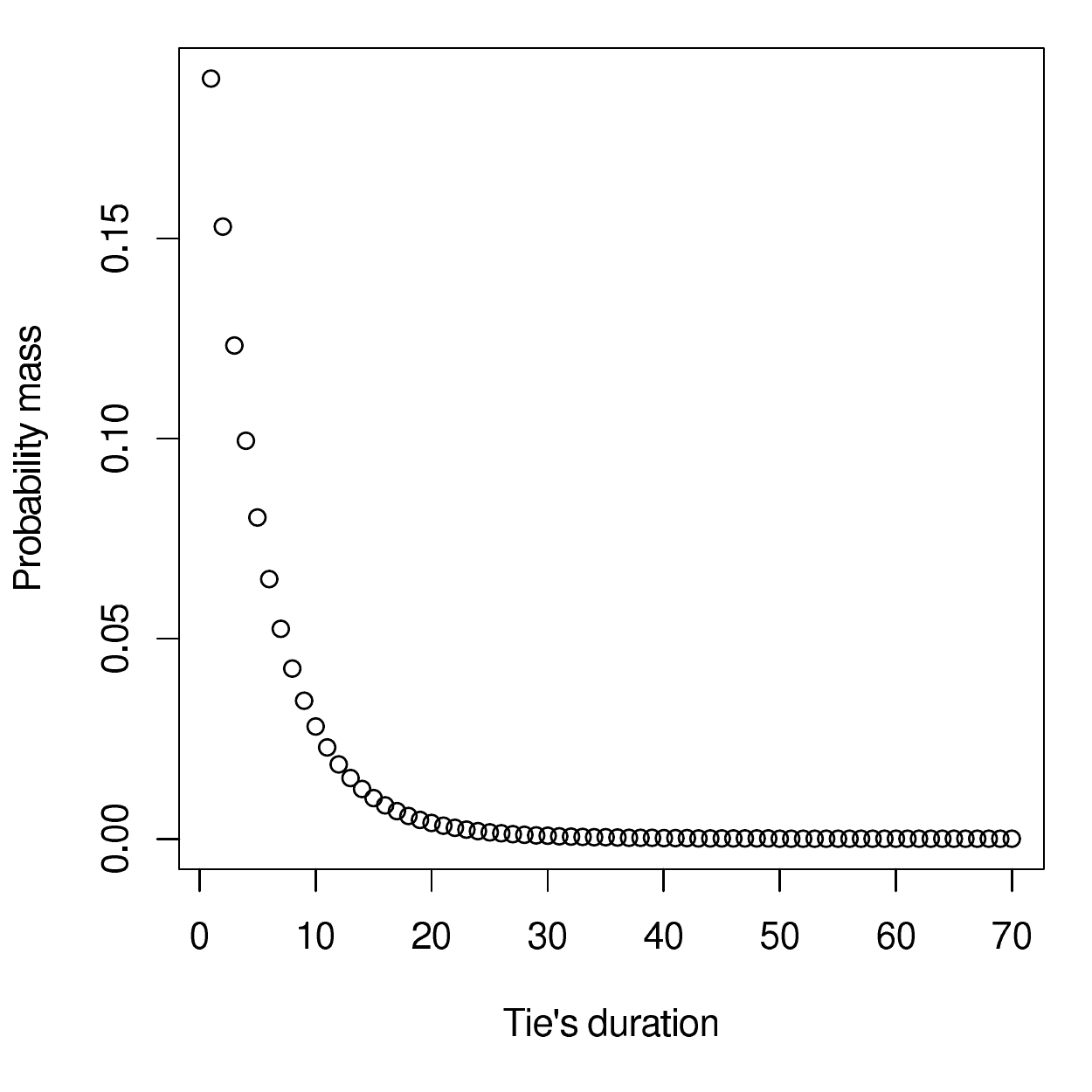}\\
        (a) hazard function&
        (b) probability mass function  \\
      \end{tabular}
    \end{center}
  }   
  \caption[Probability mass function of a mixture of two geometric
  distributions and the resulting hazard
  function]{\label{fig:2geom-hazard} Probability mass function
    of a mixture of two geometric distributions
    ($0.9\Geometric(0.2)+0.1\Geometric(0.1)$) and the resulting
    hazard function. In (a), the upper line is the hazard at a tie's
    first dissolution phase, $0.9\cdot 0.2 + 0.1\cdot 0.1$, and the
    lower line is the hazard of a tie which has persisted for a long
    time ($0.1$).}
\end{figure}

Consider a simple scenario with long-term and short-term
relationships, having $m=2$, and given that a relationship is of type
$k$, it evolves as first-order Markov, thus having a memoryless
duration distribution $\Geometric(\disspar_k)$, with $\disspar_1$
being the hazard of the short-term relationships and $\disspar_2$
being the hazard of the long-term relationships, so
$\disspar_2<\disspar_1$. Then, each type's pmf and cdf
\[f_{X_k}(x;\disspar_k)=(1-\disspar_k)^{x-1}\disspar_k,\quad F_{X_k}(x;\disspar_k)=1-(1-\disspar_k)^x,\]
leading to the marginal relationship duration distribution of
\[f_X(x;\curvparD)= \sum_{k=1}^2\bpi_k (1-\disspar_k)^{x-1}\disspar_k,\]
\[
  F_X(x;\curvparD)= \sum_{k=1}^2\bpi_k \left(1-(1-\disspar_k)^x\right)= 1- \sum_{k=1}^2\bpi_k (1-\disspar_k)^x,
\]
so
\[h_X(x;\curvparD)=\frac{\sum_{k=1}^2\bpi_k (1-\disspar_k)^{x-1}\disspar_k}{\sum_{k=1}^2\bpi_k (1-\disspar_k)^{x-1}}.\]
Then, the probability of a tie aged $x$ being preserved,
\begin{align*}
  1-h_X(x;\curvparD)&=1-\frac{\sum_{k=1}^2\bpi_k (1-\disspar_k)^{x-1}\disspar_k}{\sum_{k=1}^2\bpi_k (1-\disspar_k)^{x-1}}\\
  &=\frac{\sum_{k=1}^2\bpi_k (1-\disspar_k)^x}{\sum_{k=1}^2\bpi_k (1-\disspar_k)^{x-1}}.
\end{align*}
Initially, the probability of a tie created in the previous time step of being preserved is 
\[1-h_X(x;\curvparD)=\frac{\sum_{k=1}^2\bpi_k
  (1-\disspar_k)^1}{\sum_{k=1}^2\bpi_k
  (1-\disspar_k)^{1-1}}=1-\sum_{k=1}^2\bpi_k \disspar_k\] with hazard
$\sum_{k=1}^2\bpi_k \disspar_k$, the mean of the hazard of tie
classes, weighted by their incidence in the population. If the tie
survives, the conditional probability given its age that it was a
long-term tie in the first place increases, per Bayes's Theorem, and,
indeed,
\begin{align*}
  \lim_{x\to\infty}\left(1-h_X(x;\curvparD)\right)&=\lim_{x\to\infty}\frac{\sum_{k=1}^2\bpi_k (1-\disspar_k)^x}{\sum_{k=1}^2\bpi_k (1-\disspar_k)^{x-1}}\\
  &=\lim_{x\to\infty}\frac{\bpi_2 (1-\disspar_2)^x}{\bpi_2 (1-\disspar_2)^{x-1}}\\
  &=1-\disspar_2,
\end{align*}
so the hazard converges to $\disspar_2$, the hazard of long-term
ties. Figure~\ref{fig:2geom-hazard} for gives an example.
Notably, while this process is no longer Markovian (of any order), it
approaches a Markov process as the hazard converges.

More generally, for $m$ types of ties, the marginal duration
distribution of the mixture has
\[f_X(x;\curvparD)=\sum_{k=1}^m \bpi_k f_{\X_k}(x;\disspar),\quad
F_X(x;\curvparD)=\sum_{k=1}^m \bpi_k F_{\X_k}(x;\disspar),\]
respectively, with the discrete hazard function 
\[h_X(x;\curvparD)=\frac{f_X(x;\curvparD)}{1-F_X(x-1;\curvparD)}=\frac{\sum_{k=1}^m \bpi_k f_{\X_k}(x;\disspar)}{1-\sum_{k=1}^m \bpi_k F_{\X_k}(x-1;\disspar)}.\]

If, as in the example above, the combined hazard function converges to
some positive value that depends only on $\curvparD$, then this
duration distribution can be approximated in the STERGM framework as a
curved exponential family. Let $a_0$ be the age after which the hazard
is sufficiently close to constant. Then, setting
\[\natparD=\left(\logit(1-h_X(1;\curvparD)),\dotsc,\logit(1-h_X(a_0;\curvparD))\right),\]
and setting
\[\genstat\diss(\yD)=\left(\sum_{\pij \in \yD} \I{\age(\yD\sij)=1},\sum_{\pij \in \yD} \I{\age(\yD\sij)=2},\dotsc,\sum_{\pij \in \yD} \I{\age(\yD\sij)\ge a_0}\right),\]
leading to a change statistic 
\[\changeij\genstat(\yat{},\yat{-1})=\left(\yatij{-1}\I{\age(\yD\sij)= 1},\yatij{-1}\I{\age(\yD\sij)= 2},\dotsc,\yatij{-1}\I{\age(\yD\sij)\ge a_0}\right).\]
For any given dyad, if it has a tie at $\yatij{-1}$, all but one of
these elements will be $0$: if age $x<a_0$, then only $x$th element
will be 1. Otherwise, only $a_0$th element will be $1$, giving the
desired hazard structure.

\subsection{Hazard induced by linear age effect}
Finally, we describe a slight generalization of the piecewise-constant
hazard, in which the log-odds of a dissolution (or, equivalently of
preservation) of a tie are an affine function of the tie duration. Let
\[\genstat\diss(\yD)=\left(\abs{\yD},\sum_{\pij \in \yD} \left(\age(\yD\sij)
  \I{\age(\yD\sij)<a_0}+a_0\I{\age(\yD\sij)\ge a_0}\right)\right),\] 
with change statistic
\[\changeij\genstat(\yat{},\yat{-1})=\left(\yatij{-1},\yatij{-1}\left(\age(\yD\sij)\I{\age(\yD\sij)<a_0}+a_0\I{\age(\yD\sij)\ge a_0}\right)\right).\]
The restriction of the affine effect to the ages less than $a_0$ is to
preserve the (potentially high-order) Markov property of the process,
and to ensure that when $\curvparD_2>0$, as it would be in the
short-term--long-term scenario above, no tie would have a nonzero
probability of never dissolving.  

This dissolution statistic could be used to approximate those in
Section~\ref{sec:finite-mix-haz} more efficiently (in terms of
computing power) than the approach described in that section.

\section{Discussion}
Given that a tie does exist, we showed via a simulation study that
even in a first-order Markov model where all actors and dyads are
\emph{a priori} homogeneous, a non-geometric duration distribution ---
non-constant hazard --- can be induced by dyadic dependence in the
dissolution process.

We have also outlined several ways in which one might explicitly model
non-constant hazard durations, including piecewise-constant hazards
for situations where the duration distribution is inferred from
survival analysis, and for situations where there are substantive
reasons to model duration distribution as a mixture.

\section{Acknowlegements}
This work was supported by NIH awards P30 AI27757 and 1R01
HD068395-01, NSF award HSD07-021607, ONR award N00014-08-1-1015, NICHD
Grant 7R29HD034957, NIDA Grant 7R01DA012831, the University of
Washington Networks Project, and Portuguese Foundation for Science and
Technology Ci\^{e}ncia 2009 Program. The author would also like to
thank David Hunter and Mark Handcock, as well as the members of the
University of Washington Network Modeling Group, especially Martina
Morris and Steven Goodreau, for their helpful input and comments on
the draft.

\bibliographystyle{plainnat}
\addcontentsline{toc}{section}{References}
\bibliography{ERGM-based_models_and_inference_for_dynamic_networks}

\begin{thebibliography}{10}
\providecommand{\natexlab}[1]{#1}
\providecommand{\url}[1]{\texttt{#1}}
\expandafter\ifx\csname urlstyle\endcsname\relax
  \providecommand{\doi}[1]{doi: #1}\else
  \providecommand{\doi}{doi: \begingroup \urlstyle{rm}\Url}\fi

\bibitem[Doreian and Stokman(1997)]{doreian1997esn}
Patrick Doreian and Franz~N. Stokman, editors.
\newblock \emph{Evolution of social networks}.
\newblock Routledge, 1997.

\bibitem[Frank(1991)]{frank1991sac}
Ove Frank.
\newblock Statistical analysis of change in networks.
\newblock \emph{Statistica Neerlandica}, 45\penalty0 (3):\penalty0 283--293,
  1991.
\newblock \doi{10.1111/j.1467-9574.1991.tb01310.x}.

\bibitem[Handcock et~al.(2012)Handcock, Hunter, Butts, Goodreau, Krivitsky, and
  Morris]{handcock2012epf}
Mark~S. Handcock, David~R. Hunter, Carter~T. Butts, Steven~M. Goodreau,
  Pavel~N. Krivitsky, and Martina Morris.
\newblock \emph{\pkg{ergm}: A Package to Fit, Simulate and Diagnose
  Exponential-Family Models for Networks}.
\newblock The Statnet Project, Seattle, WA, version 3.0-1 edition, March 2012.
\newblock URL \url{http://CRAN.R-project.org/package=ergm}.
\newblock Version 3.0-1. Project home page at \url{http://www.statnet.org}.

\bibitem[Hanneke et~al.(2010)Hanneke, Fu, and Xing]{hanneke2010dtm}
Steve Hanneke, Wenjie Fu, and Eric~P. Xing.
\newblock Discrete temporal models of social networks.
\newblock \emph{Electronic Journal of Statistics}, 4:\penalty0 585--605, 2010.
\newblock \doi{10.1214/09-EJS548}.

\bibitem[Holland and Leinhardt(1977)]{holland1977dms}
Paul~W. Holland and Samuel Leinhardt.
\newblock A dynamic model for social networks.
\newblock \emph{Journal of Mathematical Sociology}, 5\penalty0 (1):\penalty0
  5--20, 1977.

\bibitem[Hunter et~al.(2008)Hunter, Handcock, Butts, Goodreau, and
  Morris]{hunter2008epf}
David~R. Hunter, Mark~S. Handcock, Carter~T. Butts, Steven~M. Goodreau, and
  Martina Morris.
\newblock \pkg{ergm}: A package to fit, simulate and diagnose
  exponential-family models for networks.
\newblock \emph{Journal of Statistical Software}, 24\penalty0 (3):\penalty0
  1--29, May 2008.
\newblock \doi{10.18637/jss.v024.i03}.

\bibitem[Krivitsky(2012)]{krivitsky2012mdn}
Pavel~N. Krivitsky.
\newblock Modeling of dynamic networks based on egocentric data with durational
  information.
\newblock Technical Report 2012-01, Pennsylvania State University Department of
  Statistics, April 2012.

\bibitem[Krivitsky and Handcock(2014)]{KrHa14s}
Pavel~N. Krivitsky and Mark~S. Handcock.
\newblock A separable model for dynamic networks.
\newblock \emph{Journal of the Royal Statistical Society, Series B},
  76\penalty0 (1):\penalty0 29--46, 2014.
\newblock \doi{10.1111/rssb.12014}.

\bibitem[Robins and Pattison(2001)]{robins2001rgm}
Garry Robins and Philippa Pattison.
\newblock Random graph models for temporal processes in social networks.
\newblock \emph{Journal of Mathematical Sociology}, 25\penalty0 (1):\penalty0
  5--41, 2001.
\newblock \doi{10.1080/0022250x.2001.9990243}.

\bibitem[Snijders(2005)]{snijders2005mln}
Tom A.~B. Snijders.
\newblock Models for longitudinal network data.
\newblock In Peter~J. Carrington, John Scott, and Stanley~S. Wasserman,
  editors, \emph{Models and methods in social network analysis}, page 215.
  Cambridge University Press, 2005.
\newblock \doi{10.1017/cbo9780511811395.011}.

\end{thebibliography}

\end{document}